# Skyrmion phase in MnSi on sapphire grown by a conventional sputtering

*Won-Young Choi,*[1] *Hyun-Woo Bang,*[1] *Seung-Hyun Chun,*[2] *Sunghun Lee,*[2,a] and *Myung-Hwa Jung*[1,a]

[1] Department of Physics, Sogang University, Seoul 04107, Korea

[2] Department of Physics, Sejong University, Seoul 05006, Korea

[a] Author to whom correspondence should be addressed: kshlee@sejong.ac.kr and mhjung@sogang.ac.kr

# ABSTRACT

Topologically protected chiral skyrmion is an intriguing spin texture, which has attracted much attention because of fundamental research and future spintronic applications. MnSi with the non-centrosymmetric structure is well-known material hosting skyrmion phase. To date, preparation of MnSi crystals has been investigated by using special instruments with ultrahigh vacuum chamber. Here, we introduce a facile way to grow MnSi films on sapphire, which is in relatively low vacuum environment of conventional magnetron sputtering. Magnetotransport properties including Hall resistivity measurements allow to confirm the existence of skyrmion phase in MnSi film. Because as-grown MnSi films on sapphire has polycrystalline nature, the emergent features of skyrmion phase are limited and complicated. However, we observed the stable skyrmion phase in a broad range of temperatures and magnetic fields, which is explained by phenomenological scaling analyses of Hall resistivities contribution. Our findings provide not only a general way to prepare the materials possessing skyrmion phase, but also insight into further research to stimulate more degrees of freedom in our inquisitiveness.

## I. INTRODUCTION

Topologically protected chiral skyrmion is a vortex-like nontrivial swirling spin texture, where magnetic spins stabilized by Dzyaloshinskii-Moriya interaction (DMI) align in non-collinear manner surrounding a sphere.[1] Large DMI is generally induced in non-centrosymmetric ferromagnets, owing to broken inversion symmetry.[2] This complex spin texture has garnered massive attention because of intriguing physical properties both for fundamental research as well as for potential of possible applications in future technology.[3,4] Compared to the magnetic domain walls, skyrmion domains exhibit stable current-driven motion at remarkably low current density, enabling low-power consumption spintronic applications.[5]

MnSi with the non-centrosymmetric B20 phase is an archetypal helimagnetic material hosting skyrmionic lattice, which has been studied theoretically and experimentally for decades.[6-10] In the skyrmionic lattice of MnSi, the spin transfer torques (STT) are observed, leading to further investigations on the injection of spin-polarized currents.[5] In especial, the skyrmion size of MnSi is ranged in ~18 nm, considered small among well-known groups with skyrmion spin texture.[11] STT tends to increase significantly with reducing the skyrmion size.[12,13] Although material parameters affect the skyrmion size, DMI coefficient and ferromagnetic exchange interaction constant mainly contribute to determine the skyrmion size.[14] In this regard, MnSi has excellent prospects to be a good candidate for applied physics.

To confirm the evident skyrmions, diverse measurement tools such as Lorentz transmission electron microscope, magnetic transmission soft X-ray microscope, magnetic force microscope, and small-angle neutron scattering are used.[15-18] Such microscopic tools allow to directly identify the skyrmionic lattice in real-space, but high-quality single crystals or epitaxial thin films are demanded, which are grown by special instruments with high-vacuum chamber. The other way to reveal the existence of skyrmions is considered to measure magnetotransport properties, topological Hall effect (THE), as shown in previous reports.[9,19-21] The skyrmion can be observed even in polycrystalline because of being a topological object,[22] which topological phase is less susceptible to impurities or crystalline nature.

Here, we report the magnetotransport properties of polycrystalline MnSi grown by conventional sputtering. We employed the X-ray diffraction (XRD) and transmission electron microscope (TEM) to identify the single phase of MnSi crystals and their crystallinity.

Temperature dependent magnetization and resistance curves showed magnetic transition at around 25 K, where magnetoresistance data also exhibited distinguishable shapes as temperatures. We successfully extracted THE signal from measured Hall resistance, and plotted contour mapping of topological Hall resistivity as a function of temperature and magnetic field. Moreover, the analysis of anomalous Hall resistivity contribution in MnSi films implied the stabilization of the skyrmion phase in a broader range of temperatures and magnetic fields, albeit impurities and defects in polycrystalline MnSi sample. Our finding provides the results that the skyrmions can be observed in polycrystalline MnSi films grown by facile and inexpensive instruments and further investigations of the similar materials possessing skyrmionic lattices can be stimulated.

## II. EXPERIMENTAL

MnSi films were deposited on Si (111) and *c*-cut sapphire ($Al_2O_3$) substrates by direct current (DC) / radio frequency (RF) magnetron sputtering with a base pressure of $1.0 \times 10^{-6}$ Torr. The MnSi films were grown at room temperature under 10 mTorr Ar pressure by co-sputtering Mn and Si target with 30 min. The DC power for Mn target was 10 to 20 W, and the RF power for Si target was 100 W. Following deposition of MnSi, as-grown MnSi was crystallized by inducing in-situ annealing treatment for 2 hours at the temperature range of 550 ~ 590 °C. The crystal phase and structure of the samples were examined by X-ray diffraction (XRD) with an X-ray source of Mo and Ag at 60 kV. The morphological characterization and chemical composition of the samples were analyzed by scanning electron microscopy (SEM), atomic force microscopy (AFM), and high-resolution transmission electron microscopy (HR-TEM) equipped with an energy dispersive spectroscopy (EDS). The magnetic and electrical properties were measured using a superconducting quantum interference device-vibrating sample magnetometer (SQUID-VSM), where the magnetic field and temperature were swept up to 50 kOe and down to 2 K, respectively.

## III. RESULTS AND DISCUSSION

The growth of MnSi films have been well described in previous reports with various methods.[2,9,21-25] However, most techniques to grow MnSi required specific facilities with ultrahigh vacuum environment, while a development for conventional magnetron sputtering with relatively low base pressure is not introduced yet. Since the lattice mismatch between out-of-plane of Si substrate and cubic MnSi structure is estimated to be around 3%,[26] we have tested to find optimal growth conditions of MnSi films on Si (111) substrates. Co-sputtering with single target of Mn and Si was employed, and growth conditions such as RF power, growth temperature, and annealing treatments were minutely controlled to grow MnSi films. (sup. Table and figure) Aguf *et al.* reported that as-deposited MnSi films was amorphous unless they were crystallized by annealing treatment.[23] Indeed, we found that initially deposited amorphous MnSi turned into crystallized MnSi phase after annealing treatment.(sup. Figure) Most results using Si (111) substrate, however, showed that mixed phases of MnSi and $Mn_5Si_3$ were observed by XRD measurements. For this reason, Si (111) substrates were replaced into $Al_2O_3$ substrates due to low lattice mismatch.

Figure 1 presents the XRD patterns of MnSi films grown on Si (black solid line) and $Al_2O_3$ (blue and red solid lines), where MnSi on Si (111) and on $Al_2O_3$ #1 were deposited with same growth conditions (15 W for Mn power, 100 W for Si, 590 ℃ annealing treatment). The asterisk in the Figure indicates $Mn_5Si_3$ (ICSD card no. 04-003-4114) phase. Substituted $Al_2O_3$ for Si substrates, the peaks related with $Mn_5Si_3$ phase were suppressed, and furthermore MnSi on $Al_2O_3$ #2 sample, for which Mn power and annealing temperature deceased to 10 W and 550 ℃, respectively, showed only MnSi (ICSD card no. 04-004-7568) peaks.

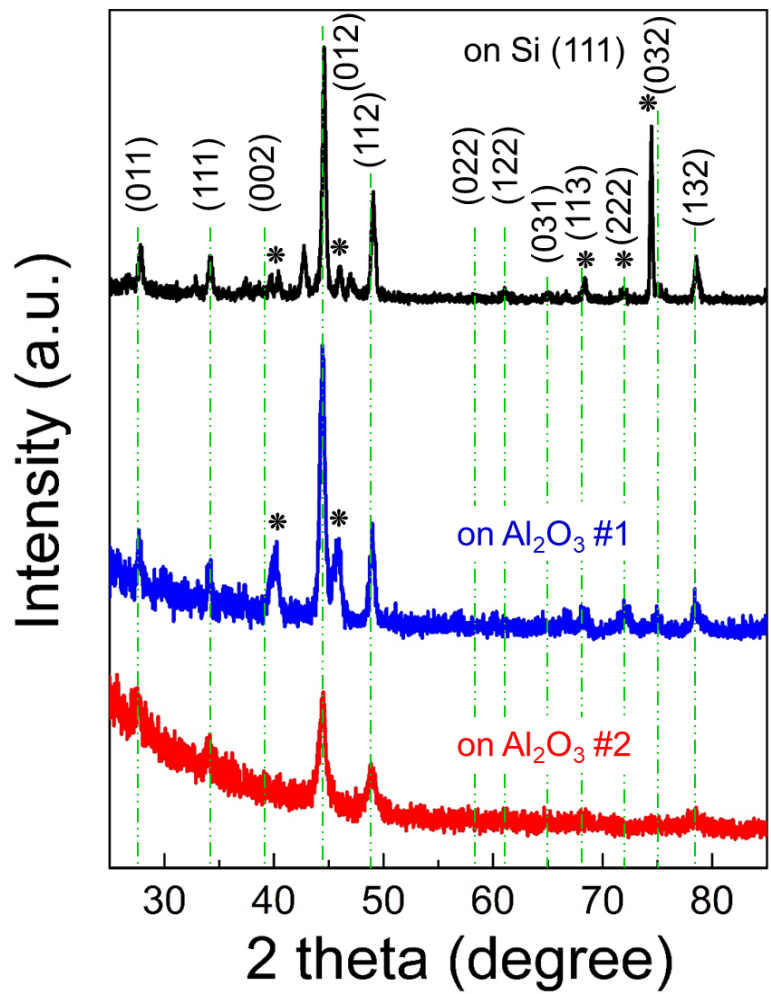

**Figure 1.** XRD pattern of MnSi films on Si [(111), black solid line] and sapphire [(Al$_2$O$_3$), blue and red solid lines] substrate. All the peaks are indexed to the cubic B20-type MnSi phase, marked with green dotted lines. The asterisks in black and blue solid lines indicate the peaks from Mn$_5$Si$_3$ phase.

Although as-grown MnSi on Al$_2$O$_3$ samples shows somewhat defective surface, highly uniform and low uneven surface is observed, as shown in SEM image of Figure 2(a) and AFM topographic image of Figure 2(b). In the 15×15 μm scale of AFM image, the root mean squared (RMS) roughness is measured to be under 1 nm. To characterize detailed structure and chemical composition, the cross-sectional TEM analyses of as-grown MnSi on Al$_2$O$_3$ sample are carried out. Figure 2(c) shows the representative cross-sectional TEM image of MnSi on Al$_2$O$_3$ sample at the interfacial region. Note that no stacking faults nor significant defects are observed. When one grows MnSi films by conventional sputtering in relatively low vacuum chamber, we found that MnSi did not follow epitaxial growth to the preferred direction of the surface of substrates, determined by structural parameters such as lattice mismatch and chemical bonding. Our MnSi samples grown on $Al_2O_3$ have polycrystalline nature, confirmed by XRD patterns (Figure 1) and fast Fourier transform (FFT) of TEM image [inset of Figure 2(c)]. We have examined chemical composition of as-grown MnSi films. As seen in TEM-EDS mapping of Figure 2(d), the presence of only Mn and Si elements is detected at several different regions, and the atomic ratio between Mn and Si with Mn/Si = 1:1.1 is estimated. We tested growth rate of MnSi films by controlling growth time. The thickness of as-grown MnSi films showed a linear behavior for the growth time (see sup. Info.)

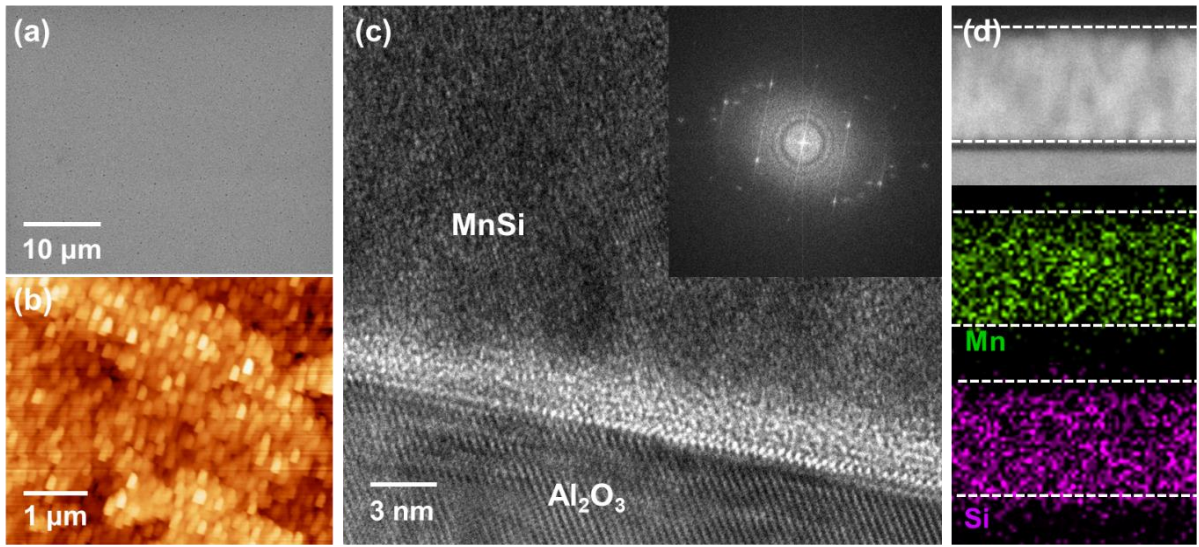

**Figure 2.** Morphological and structural characterization of MnSi films grown on $Al_2O_3$ substrate. (a) SEM image of the as-grown MnSi film. (b) AFM topographic image corresponding to Figure 2(a). RMS roughness is estimated to be under 1 nm. (c) Representative HR-TEM image of MnSi film grown on sapphire. Inset: FFT from selected area of MnSi in HR-TEM image. (d) Elemental mapping of the EDS of the cross-sectional MnSi film.

Figure 3(a) shows the temperature dependence of magnetization of MnSi on $Al_2O_3$ (thickness; 150 nm) with an out-of-plane magnetic field of 1 kOe. The magnetization dropped significantly at temperatures above 30 K, indicating transition temperature ($T_C$), similar to the bulk MnSi.[27,28] The resistance depending on the temperature exhibited metallic behavior above transition temperature. Below $T_C$, the resistance tended to decrease with $T^2$ dependence as decreasing temperatures, owing to coupling of the charge carriers to spin fluctuations in helimagnetic phase.[29] As seen in inset of Figure 3(b), the derivative of resistance versus temperature highlighted $T_C$ of MnSi films, around 25 K. The slight difference of $T_C$ from M-T curve might be due to polycrystalline nature. Also, polycrystal and defects on the surface give rise to the low residual resistance ratio, [R(300 K) / R(5 K)] ~ 1.7.

Figure 3(c) shows the magnetoresistance for the magnetic fields perpendicular to the sample at different temperatures of 2 K, 25 K, and 50 K. As we discussed above, since as-grown MnSi films have polycrystalline nature, the magnetic phase transition from the magnetoresistance data revealed unclearly. In low magnetic fields, however, a behavior of the magnetoresistance at different temperatures exhibited distinguishable feature. As the temperature increases, shape of the magnetoresistance in the vicinity of zero magnetic field changes from flat to sharp and broad peak.

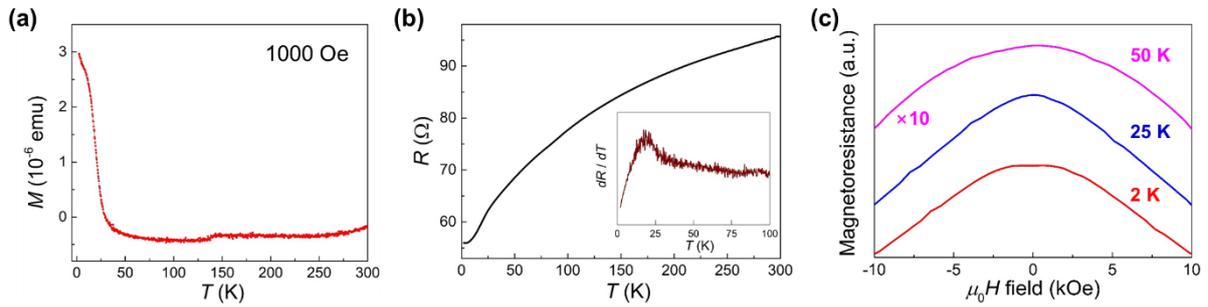

**Figure 3.** (a) Field-cooled magnetization as a function of temperature for 150 nm thick MnSi film in an external magnetic field of 1 kOe. (b) Zero-field longitudinal resistance as a function of temperature. Inset: derivative of the resistance vs temperature highlighting the anomaly of magnetic transition. (c) Perpendicular magnetoresistance at 2, 25, and 50 K. For clarity, the arbitrary offsets are added, and the magnetoresistance measured at 50 K is magnified as 10 times.

Spin-chirality-driven Hall effect, THE can be induced by DMI arising out of strong spin-orbit coupling and non-centrosymmetric B20 crystal structure,[30] considered as a hallmark for the existence of skyrmion phase. We have performed Hall resistivity measurement to observe abnormal resistivity related with THE. The total Hall resistivity can be expressed as a combination of three components:

$$\rho_{Hall} = \rho_{normal} + \rho_{AHE} + \rho_{THE} = R_0 H + (\alpha \rho_{xx0} + \beta \rho_{xx0}^2 + b\rho_{xx}^2)M + n_{Skx} P R_{TH} B_{eff},$$

where $\rho_{normal}$, $\rho_{AHE}$, and $\rho_{THE}$ are the normal, anomalous, and topological Hall resistivities, respectively. $R_0$ is normal Hall coefficient, and $\alpha$, $\beta$, and $b$ are constants corresponding to the skew scattering, side jump, and intrinsic contributions to the anomalous Hall resistivity. Also, $n_{Skx}$ is the relative skyrmion density, $P$ is the polarization of the conduction electrons, $R_{TH}$ is the topological Hall coefficient, and $B_{eff}$ is the effective magnetic field derived from the real-space Berry phase.[20,31] Topological Hall contribution can be extracted by subtracting the normal and anomalous Hall resistivity terms from the measured total Hall resistivity.

Figure 4(a) shows the extracted THE signal at 10 K as the blue curve, including normal (green line) and anomalous (red curve) Hall resistivities. Note that the negative slope of $\rho_{normal}$ indicates $p$-type charge carriers, and $\rho_{AHE}$ is negative, consistent with those of MnSi bulk,[32] thin film,[9] and nanowire.[20] The corresponding $\rho_{normal}$ and $\rho_{AHE}$ derived from the magnetotransport and SQUID data analysis. THE contributed resistivity depending on the temperature is displayed in Figure 4(b). Interestingly, the sign of $\rho_{THE}$ flips at the border of 25 K, where magnetic transition is expected. The sign of $\rho_{THE}$ is very sensitive to the spin polarization of charge carriers. In the band structure of MnSi, $d$ electrons affect the density of states near Fermi level, but rather localized, while $s$ band that contributed meagerly in band structure is of itinerant,[32] allowing the spin polarization to be delicate. In addition, since the spin polarization can be changed by external factors such as tensile strain and crystal purity with temperature,[9] the discrepancy of the sign of $\rho_{THE}$ in our polycrystalline MnSi sample is reasonable. Figure 4(c) presents the contour mapping of $\rho_{THE}$ as a function of magnetic field and temperature. While skyrmion phase in bulk MnSi is observed in narrow temperature range close to the magnetic transition temperature, non-zero $\rho_{THE}$ is collected from 2 K to 40 K regardless the sign. The absolute value of $\rho_{THE}$ has a maximum value of 36 nΩ.cm at 10 K and 4 kOe, larger than thin films grown by MBE (10 nΩ.cm),[9] bulk (4.5 nΩ.cm),[33] and nanowire (15 nΩ.cm)[20] but similar to the thin films grown by off-axis magnetron sputtering with ultrahigh vacuum chamber.[25]

$\rho_{AHE}$ consists of three components; skew scattering, side jump, and intrinsic contribution. Implication in the scaling of anomalous Hall contribution is that $\rho_{AHE}$, is proportional to the intrinsic contribution, $\rho_{xx}^2$, associated with momentum-space Berry-phase.[34] In Figure 4(d), we plot the $\rho_{AHE}$ against $\rho_{xx}^2$ at 20 kOe, showing obvious deviation from linear dependence. The breakdown of the scaling is suggested that anomalous Hall effect is relevant to extrinsic skew scattering and side jump contributions caused by impurities and defects in our polycrystalline MnSi sample, remaining the stabilization of the skyrmion phase in a broader range of temperatures and magnetic fields.

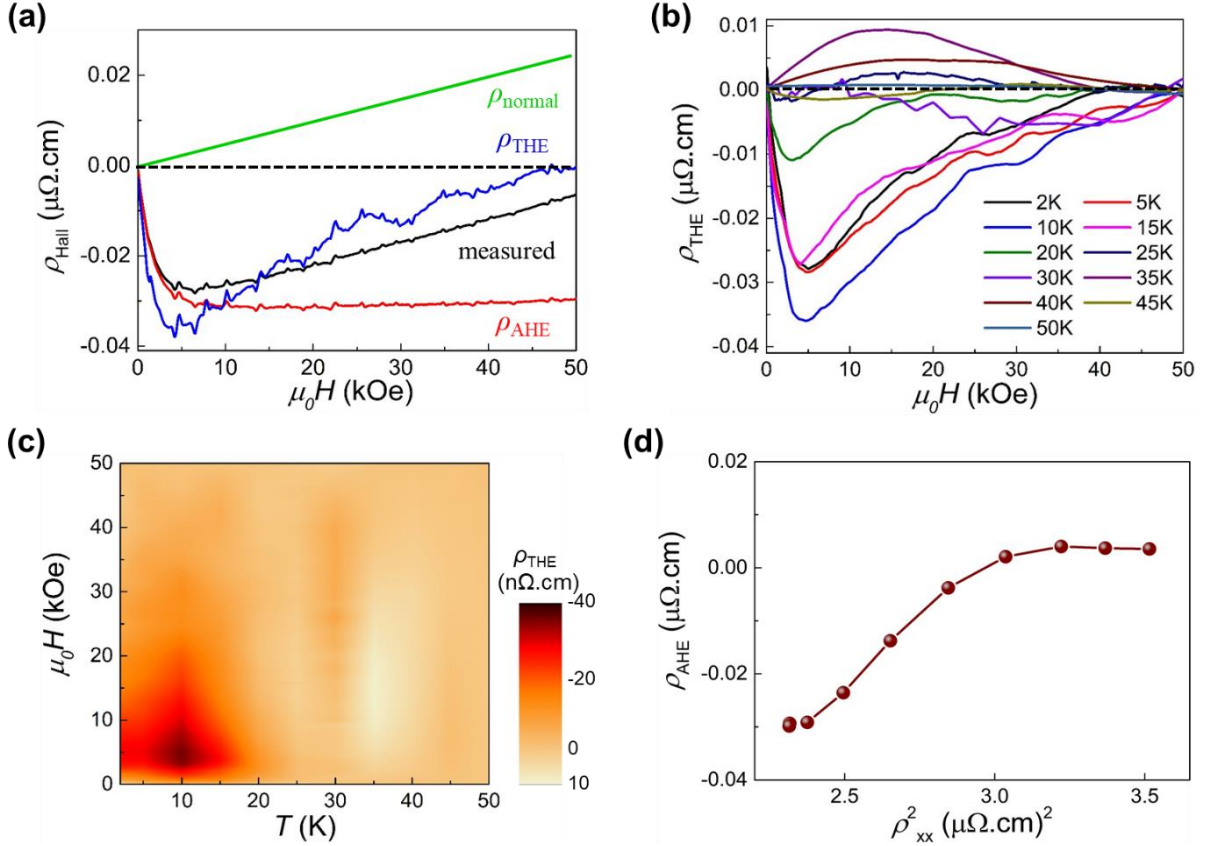

**Figure 4.** (a) The representative Hall resistivity curve at 10 K. THE signal (blue curve) is extracted by subtracting normal (green line) and anomalous Hall signal (red curve) from total measured Hall resistivity (black curve). (b) topological Hall resistivities at various temperatures, extracted using same procedure detailed in the text. (c) The contour mapping of THE signal as a function of the magnetic field and temperature, constructed by interpolation of topological Hall resistivity between temperatures. (d) Anomalous Hall resistivity as a function of squared longitudinal magnetoresistivity below the temperature, where topological Hall resistivity is not zero.

## IV. CONCLUSION

In summary, we demonstrated a method to grow MnSi films on $Al_2O_3$ by conventional magnetron sputtering with relatively low vacuum chamber. The spectroscopic and morphological analyses confirmed that as-deposited MnSi films had polycrystalline nature with highly uniform and low rough surface. The transport properties the intrinsic characteristics of MnSi, though magnetic transition temperature is slightly lower than that of previous results. More importantly, we observed stable skyrmion phase in a broad range of temperatures and magnetic fields even in our polycrystalline MnSi films, attributed to the complicated implication of Hall resistivities contribution. This work opens up the opportunity for extensive investigation on the materials possessing skyrmion phase, away from the burden to prepare single crystals or epitaxial thin films

**SUPPLEMENTARY MATERIAL**

See the supplementary material for summary of growth test of MnSi film on $Al_2O_3$ substrate, XRD pattern evolution with respect to the annealing treatment, and growth rate data.

**ACKNOWLEDGMENTS**

This work was supported by the National Research Foundation of Korea (NRF) grant funded by the Korea government (MSIT) (Nos. 2016M3A7B4910400, 2016R1E1A1A01942649, 2017R1A2B3007918, 2018R1A5A6075964, 2018K1A4A3A01064272, 2018R1D1A1B07048109).